# Low temperature mixed spin state of $Co^{3+}$ in $LaCoO_3$ evidenced from Jahn-Teller lattice distortions

V. Gnezdilov,[1] K.-Y. Choi,[2] Yu. Pashkevich,[3] P. Lemmens,[4]
S. Shiryaev,[5] G. Bychkov,[5] S. Barilo,[5] V. Fomin[1], and A.V. Yeremenko[1]

[1] *B.I. Verkin Inst. for Low Temp. Physics NASU, 61164 Kharkov, Ukraine*
[2] *Institute for Material Research, Tohoku University, Katahira 2-1-1, Sendai 980-8577, Japan*
[3] *A.A. Galkin Donetsk Phystech NASU, 83114 Donetsk, Ukraine*
[4] *Institute for Physics of Condensed Matter, TU Braunschweig, D-38106 Braunschweig, Germany*
[5] *Institute of Physics of Solids & Semiconductors, Academy of Sciences, 220072 Minsk, Belarus*
E-mail: gnezdilov@ilt.kharkov.ua

One- and multi-phonon excitations of the single crystalline $LaCoO_3$ were studied using Raman spectroscopy in the temperature region of 5 K – 300 K. First-order Raman spectra show a larger number of phonon modes than allowed for the rhombohedral structure. Additional phonon modes are interpreted in terms of activated modes due to lattice distortions, arising from the Jahn-Teller (JT) activity of the intermediate-spin (IS) state of $Co^{3+}$ ions. In particular, the 608-$cm^{-1}$ stretching-type mode shows anomalous behavior in peak energy and scattering intensity as a function of temperature. The anomalous temperature dependence of the second-order phonon excitations spectra is in accordance with the Franck-Condon mechanism that is characteristic for a JT orbital order.

I. Introduction

Lanthanum cobalt oxide $LaCoO_3$ is one of a series of cobaltites with the chemical formula $RCoO_3$, where *R* is a rare-earth element or Y. From earlier studies it has been known, that $LaCoO_3$ undergoes successive spin-state transitions as a function of temperature[1-10]. The first spin transition takes place near 100 K from a low-spin (LS, $t_{2g}^6 e_g^0$, $S = 0$) state to an intermediate-spin (IS, $t_{2g}^5 e_g^1$, $S = 1$) state. The second one occurs near 500 K from an IS state to a high-spin (HS, $t_{2g}^4 e_g^2$, $S = 2$) state of $Co^{3+}$ ions. Here note that due to the partially filled $e_g$ level, the IS state is Jahn-Teller (JT) active.

Although the spin-states of cobalt ions and the underlying nature of the transitions between them have been investigated for over 50 years, this topic has recently gained new interest. In midst of the research activity, the role of the crystal structure of $LaCoO_3$ remains rather controversial. Another issue concerns a possible orbital ordering in the IS state. Raman spectroscopy is in general a useful tool for such and related questions.

As mentioned in our previous work[11], Raman spectra of $LaCoO_3$ consist of a larger number of one-phonon modes than what is expected for the rhombohedral structure. In this paper we report on polarized Raman spectra of single crystalline $LaCoO_3$ measured in the temperature range of 5 – 300 K. Additional modes observed in whole temperature range are attributed to phonon scattering that are activated in the Raman scattering (RS) process by structural distortions related to a mixed LS/IS state. Detailed temperature dependent measurements allowed us to identify a coupling of the 608-$cm^{-1}$ optic phonon mode to JT distortions of $CoO_6$ octahedra as well as the influence of the JT orbital ordering on multiphonon scatterings. Moreover, anomalies in peak energy and relative intensity of the corresponding modes give evidence for the presence of the IS $Co^{3+}$ ions together with the JT distorted $CoO_6$ octahedra even at lowest temperature of the previously assumed pure LS state.

II. Experimental details

Single crystals of $LaCoO_3$ were grown using an anodic electro-deposition technique. In particular, McCarrol et. al approach[12] was modified to use seeded flux melt growth based on $Cs_2MoO_4$-$MoO_3$ mixture in the ratio 2.2 : 1 as solvent.[13] Appropriate amount of solvent was added into a 100 $cm^3$ platinum crucible containing the mixture to grow these single crystals with a seed served as an anode at ~ 950 – 1000 °C under current density in the range 0.5 – 0.7 mA/cm$^2$. Simultaneously, the crucible serves as a cathode of the electrochemical cell.

Raman scattering measurements were carried out in quasi-backscattering geometry using 514.5 nm line of an argon laser. The incident laser beam of 10 mW power was focused onto a 0.1 mm spot of the mirror-like chemically etched surface of the as grown crystal. The sample was mounted on the holder of a He-gas flow cryostat using silver glue. The scattering light was analyzed with a DILOR XY triple spectrometer combined with a nitrogen-cooled CCD detector. Provided the naturally grown surfaces of the perovskite-like crystals are the quasi-cubic ones, the measurements were performed in the *xx*, *xz*, *x'x'*, and *x'z'* scattering configurations, where *x*, *z*, *x'*, and *z'* are the [100], [001], [101], and [$\bar{1}$01] quasi-cubic directions, respectively.

III. Results and discussion

In certain cases, it is extremely difficult to identify the exact crystal symmetry. In all previous studies, based on powder X-ray and neutron diffraction measurements, the crystal structure of LaCoO$_3$ was interpreted as rhombohedral without any structural transitions in the temperature interval of 4.2 – 1248 K.[14-17] The rhombohedral $R\bar{3}c$ structure can be obtained from the simple cubic perovskite (P*m*3*m*) by a rotation of the adjacent CoO$_6$ octahedra in opposite directions around the cubic [111] direction. For the rhombohedral structure, the factor group analysis yields five Raman-active modes ($A_{1g} + 4E_g$) out of the total 20 Γ-point phonon modes.

In contrast to the common interpretation of the LaCoO$_3$ crystal symmetry as $R\bar{3}c$, a recent powder and single crystal X-ray diffraction study[18] as well as thermal expansion measurements[19] and neutron pair distribution function analysis[20] provide evidence for the presence of monoclinic distortions in LaCoO$_3$. The monoclinic distorted phase is proposed to be due to the strain caused by a cooperative JT effect.[18] The averaged structure was found to be *I*2/*a* with three unequal Co-O bond lengths in the LS/IS state: one short, one long, and one medium length bond.[18] The long and short Co-O distances correspond to the bonds in the *ab* plane while the medium Co-O distance is the out-of-plane bond all in *I*2/*a* setting. The number of structural items in the monoclinic phase remains the same as in the rhombohedric one and the inversion symmetry also preserves.

The polarized Raman spectra of LaCoO$_3$ measured on a quasi-cubic (001) surface at 5 K are shown in Fig. 1. The inset of Fig. 1 displays the temperature dependence of the magnetic susceptibility of the studied sample. The sharpness of the observed phonon modes in Fig. 1 and an agreement of our susceptibility data with the previously reported measurements[15,21,22] indicate a high quality of our single crystals. The total number of the observed modes is at least three times larger than what is expected for the $R\bar{3}c$ structure. Among them three pronounced peaks are seen at 562, 657, and 785 cm$^{-1}$. A closer inspection reveals that the most intense peak at 657 cm$^{-1}$ in the *xx* and *x'x'* polarizations is composed of two additional peaks at 608 and 701 cm$^{-1}$, which are well visible in the *x'z'* polarization. In addition, peaks at 48, 82, 122, 138, 167, 196, 247, 286, 340, 371, 405, and 432 cm$^{-1}$ with different polarizations as well as a maximum centered at ~485 cm$^{-1}$ with a three-peak structure are observed.

The Raman spectra of LaCoO$_3$ are, however, somewhat surprising in the sense that they have no similarity to those of the isostructural rhombohedral compounds such as LaMnO$_{3+\delta}$,[23-26] La$_{1-x}$A$_x$MnO$_3$,[23,24] LaAlO$_3$.[26] Even, they differ from the spectra of LaCoO$_3$ reported recently by Ishikawa *et al.*.[27] However, the systematic evolution of the spectra from pure LaMnO$_3$ to pure LaCoO$_3$ in the series of La$_{1-x}$Co$_x$MnO$_3$[11] builds confidence of our data.

The exact assignment of the phonon modes, relying on their polarization dependence, is complicated due to the twinning of the crystals which is common for perovskites. However, compared to lattice-dynamical calculations and experimental data of Ref.26 for rhombohedral LaAlO$_3$, one can safely assign the mode at 138 cm$^{-1}$ to a rotation of the oxygen octahedra around the hexagonal [001] direction, the peak at 167 cm$^{-1}$ to pure La vibration in the hexagonal (001) plane, and the peaks at 485 and 657 cm$^{-1}$ to the internal (bending- and stretching-like, respectively) vibrations of the CoO$_6$ octahedra. Note that in comparison to our data, the out-of-phase stretching mode of the isostructural ABO$_3$ compounds is very weak.

In principle, a reduction of the crystal symmetry from $R\bar{3}c$ to $I2/a$ might explain the appearance of new phonon modes in the Raman spectra. Three $A_{2g}$ modes of rhombohedric phase become Raman active $3B_g$ modes of monoclinic phase while every $E_g$ mode should be splited into $A_g + B_g$. Nonetheless, twelve ($5A_g+7B_g$) Raman active modes expected for $I2/a$ group are not enough to explain the number of the peaks observed in our Raman experiments.

Another source of new phonon modes can be attributed to local lattice distortions associated with the thermally induced local IS state of $Co^{3+}$ ions in a matrix of $Co^{3+}$ ions of the LS state. Noticeably, an infrared spectroscopy study of $LaCoO_3$ has shown anomalies in the phonon spectra, which are ascribed to local lattice distortions.[22] It is well known that in such cases the selection rules for an average structure may be violated. More specific, the short range local lattice distortions break a rotational symmetry and can lead to the appearance of new phonon modes originating from the Brillouin zone boundaries, which are forbidden in the average, global crystal symmetry. In this case, the activated phonon modes should be weak. In our case week phonon peaks are observed in the frequency regime of 200-450 $cm^{-1}$ (see Fig.1).

In the case of phase separation sets of phonons characteristic for different phases must simultaneously present in the Raman spectra of the sample. As an example of such phase coexistence is the observation of JT distorted and non-JT distorted regions in the manganite samples.[28,29] Note that Raman intensities for each phase in this case will be proportional to the relative phase volume in the sample.

To detect possible structure changes through the LS-IS state transition, we have examined in detail the temperature dependence of optical phonon spectra (Fig. 2). At first glance, there seems to be no appreciable changes of the spectra in the temperature interval of 5 K – 300 K. To extract more detailed information, we performed a dispersion analysis of the spectra in the frequency region of 500 – 850 $cm^{-1}$, where the most intense peaks are observed. Lorentzian profiles were used to fit the spectra as shown in Fig. 3. The temperature dependence of phonon frequencies is plotted in Fig. 4. With increasing temperature the 657- and 701-$cm^{-1}$ modes soften roughly by 8-11 $cm^{-1}$. In contrast, the 608-$cm^{-1}$ mode shows an anomalous behavior; upon heating up to 130 K, first it hardens by ~10 $cm^{-1}$ and then it softens by ~12 $cm^{-1}$. The 468-, 483-, and 494-$cm^{-1}$ modes show a monotonic decrease of frequency by 7-11 $cm^{-1}$ with increasing temperature (not shown here). In order to explain possible origins of the exceptional behavior seen in the 608-$cm^{-1}$ mode we estimate the phonon frequency shift due to a thermal expansion using a Grüneisen law, $\Delta\omega/\omega_i = -v_i \Delta V/V$ (V is the unit cell volume and $v_i$ is the Grüneisen parameter for the i-th phonon mode). Lattice parameters are taken from Refs. 15 and 17 for a calculation of the unit cell volume change. The results are displayed by the dotted lines together with the raw data in Fig. 4. There is a good agreement between the estimated and the observed behavior for the 657- and 701-$cm^{-1}$ phonon modes. However, a strong deviation shows up for the 608-$cm^{-1}$ phonon mode in the temperature range of 5 - 130 K. Noticeably, in the corresponding temperature regime the magnetic susceptibility exhibits a rapid drop which connected with gradual decrease of IS state population under temperature decrease.[15,21,22] This can indicate a strong coupling of the 608-$cm^{-1}$ optic phonon mode with the IS spin state of the cobalt ions.

To get more insight, we will examine the temperature dependence of the phonon integrated intensity. In Fig. 5 the relative integrated intensities of the respective modes, $I_i/I_{tot}$, are plotted. The intensity of the 657- and 701-$cm^{-1}$ modes decreases with increasing temperature while the intensity of the 608-$cm^{-1}$ mode increases. The behavior of the other strong peaks in the frequency region of 500 – 800 $cm^{-1}$, the phonon modes at 562 and 785 $cm^{-1}$, exhibits a temperature dependence of the integrated intensity similar to the mode at 608 and 657 $cm^{-1}$, respectively (not shown here). The phonon modes related to the bending-type vibrations are split into three peaks as the stretching-type modes are. The intensity of these three peaks as a function of temperature is displayed in Fig. 6a. Figure 6b displays an example of the spectra fitting in the frequency region of the bending-type vibrations. Note that the intensity of the 494-$cm^{-1}$ modes also demonstrates anomalous temperature dependence similar to the 608-$cm^{-1}$ mode. In contrast to the 608-$cm^{-1}$ mode, however, the 494-$cm^{-1}$ mode shows no substantial softening of its peak energy upon cooling below 130 K (not shown here).

The contrasting behavior of the observed phonon's intensity and frequency (see Figs. 4-6) can be explained by the following scenario. One can suppose that oxygen octahedra with $Co^{3+}$ ions in different spin states possess a different Raman response (Raman tensor) because of difference in $Co^{3+}$ radii and, more important, due to an increase of covalency of the oxygen-metal bonds in the IS state.[30] The change of the covalency has also a strong impact on the lattice dynamics. However, the effect of all these circumstances is very selective to the symmetry of a given vibration of the oxygen octahedra. For instance, it is clear that the frequency of the stretching-like mode must increase upon an increase of covalency while the frequency of the bending-like mode is not so sensitive to the $p$-$d$ hybridization. Applying these considerations to the 494- and 608-$cm^{-1}$ modes, one can assign them to the bending and stretching type of oxygen octahedra vibration, respectively, which both originate from $E_g$ modes of the rhombohedric phase. The strongest evidence of this assignment comes from their intensities as a function of temperature that has to reflect the increasing number of JT distorted $CoO_6$ octahedra ($Co^{3+}$ in the IS state) due to the temperature-induced increasing population of this spin state. This is indeed observed in our spectra (see Figs.5,6). Note that this assignment also is in accordance with the lattice dynamical calculations of the $E_g$ stretching- and bending-like modes in $LaAlO_3$.[26]

Then we arrive at the strongest phonon line at 657 $cm^{-1}$. Assuming that this phonon mode is the $E_g$ stretching vibration in the rhombohedric phase one can explain the unusual temperature dependence of its integrated intensity. Indeed, in this case the contribution to the scattering intensity comes from the population of undistorted $CoO_6$ octahedra ($Co^{3+}$ in the LS state).

Based on above arguments we can finally assign lines at 468 and 657 $cm^{-1}$ to $E_g$ bending and stretching vibrations in the non-JT distorted phase, respectively. Lines at 483, 494 $cm^{-1}$ can be assigned to stretching- and lines at 608, 701 $cm^{-1}$ to bending-type vibrations of $A_g$ or $B_g$ symmetry in the JT distorted phase. An assignment of all lines in the spectra of $LaCoO_3$ need in more detailed study. Herewith, it is necessary to take into account possible more complicated scenario of crystal structure changing via temperature starting from single local distortions at low temperatures to macroscopic phase separation at higher temperatures.

As was mentioned above, integrated intensities of phonon modes associated with different phases must follow the relative volume of the corresponding phase. Figure 5 presents the scaled populations, $x_s$ (S = 0, 1), for the $Co^{3+}$ ions in LS and IS states. We have used the model of Ref. 15 and the same parameters for the calculation of $x_s$. The correlation of the relative integrated intensities between the 608- and 657-$cm^{-1}$ modes and the populations $x_0$ and $x_1$, respectively, are not unreasonable except the temperature regime below 50 K for both modes. Here, the 657-$cm^{-1}$ mode does not reach its expected maximum. In contrast, the 608-$cm^{-1}$ mode has a finite intensity even at lowest temperature. This implies that a pure LS state is never achieved even at lowest temperature, that is, a small amount of IS $Co^{3+}$ ions are still present for temperatures below 50 K. A similar conjecture has been made in Ref. 21. Furthermore, our study is consistent with the analysis of thermal expansion and magnetization measurements[15] which also show the absence of a pure spin state at finite temperature. An evidence for a monoclinic distorted structure of $LaCoO_3$ in the temperature interval 20 - 300 K was presented in Ref. 18. It was concluded that the monoclinic distortion state is brought about by a cooperative JT effect which triggers the long range orbital ordering of the $e_g$ orbitals. In Ref.27 it was concluded that the magnetic state in $LaCoO_3$ changes from the LS to the mixed state with the thermally excited IS at higher (~ 50 K) temperature, taking no consideration on the fact that the $A_{2g}$ modes, which are Raman inactive in the rhombohedral $R\bar{3}c$ phase (those modes are active in the monoclinic $I2/a$ phase), were observed in the spectra in a whole temperature range from 5 to 300 K. We do not exclude that the observation of the extra phonon lines at low temperatures in our experiments might be due the surface effects: local distortions or presence of $Co^{3+}$ ions with IS state near the surface of the sample.

We will turn now to the high-frequency Raman scattering response. In the $LaCoO_3$ crystal the one-phonon peak at 785 $cm^{-1}$ overlaps with a broad maximum centered at 895 $cm^{-1}$ (0.11 eV) (Fig. 7). We will discuss now its possible origin. It might originate from a photoionization of small polarons, i.e., an electronic transition from a band of localized small polaron states to a conduction band. A Raman scattering study of the paramagnetic phase of $A_{1-x}A'_xMnO_3$ system[31] shows such a maximum at ~1100 $cm^{-1}$ (0.14 eV) with nearly the same spectral shape as well as with a comparable temperature dependence of the spectrum. However, this origin is unlikely in our case as undoped $LaCoO_3$ has no holes which can form polarons. Furthermore, infrared reflectivity measurements[32] unveil that electron-phonon coupling is not large with respect to any particular phonon and that rather large polaronic states would be expected. The broad maximum at 895 $cm^{-1}$ could be considered also as second-order scattering of the one-phonon modes at 468, 483, and 494 $cm^{-1}$. Finally, the observed maximum can arise from an electronic transition through the charge gap. The energy of the observed excitation is very close to the calculated (0.2 eV)[33] and the observed values of the charge gap ($\approx$0.1 eV)[34] by optical conductivity. Most probably, the observed maximum is a mixture of electronic excitations through the charge gap and multiphonon scattering.

Finally, we will focus on the higher frequency region of the spectra where multiphonon scattering is observed (Fig. 8). Our measurements show three features at about 1215, 1305, and 1564 cm$^{-1}$, which correspond approximately to twice the energy of the one phonon modes at 608, 657, and 785 cm$^{-1}$, respectively. However, not all phonon modes show up as higher-order scattering. For instance, we find no evidence for the presence of second-order signal of the strong polarization-dependent mode at 562 and 701 cm$^{-1}$. A quantitative characterization of the second–order Raman response is summarized in Fig. 9. The ratio of a second- to first-order integrated Raman intensities, that is, $I^{(2)}/I^{(1)}$ decreases gradually with increasing temperature for the mode at 608-cm$^{-1}$. Such a behavior is rather expected because both the IS and the LS sites have to contribute to the two–phonon Raman process. In contrast, the one-phonon intensity depends just on the population of IS state. However, this intensity develops much faster than the ratio of $(x_0+x_1)/x_1$ (this ratio can be easily depicted from the Ref. 15) does. A nearly temperature-independent ratio of $I^{(2)}/I^{(1)}$ seen for the 657- and 785-cm$^{-1}$ modes as well as for their corresponding overtones at 1305 and 1564 cm$^{-1}$ is observed (see Fig. 9). Such a difference directs us to emphasize two different mechanisms leading to the corresponding two-phonon scattering intensity.

Recent theoretical[35,36] and experimental[37,38] studies of multiphonon scattering in manganites reveal that in an orbital ordered state the conventional two-phonon Raman process is superimposed by two-phonon scattering induced by the Franck-Condon (FC) mechanism. These two processes depend in different order of perturbation theory on electron-phonon interaction constant. They are also selective to the symmetry of vibrations. In the case of a dominant FC contribution the ratio, $I^{(2)}/I^{(1)}$, is expected to be temperature independent.[36,37] Thus, the interrelation between first- and second-order Raman scattering of the 657- and 785-cm$^{-1}$ modes, shown in Fig.9, strongly indicates the presence of the FC mechanism, i.e. orbital order[18,27] in LaCoO$_3$ at least on a short range scale and even at low temperatures. Furthermore, on heating, the ratio $I^{(2)}/I^{(1)}$ demonstrates a small increase, which is in accordance with an increasing population of the IS state.

IV. Conclusion

In summary, we have reported first- and second-order Raman scattering on the single crystalline LaCoO$_3$. The first-order spectra exhibit a larger number of phonon modes than allowed for rhombohedral ($R\bar{3}c$) symmetry. Additional modes are identified as activated oxygen octahedra modes due to JT distortions. The temperature dependence of the breathing - and stretching-type phonon modes on cooling suggests the presence of Co$^{3+}$ ions in the intermediate spin state, even at lowest temperatures. In the high-frequency region an anomalous temperature dependence is observed for the two-phonon scattering at 1305, and 1564 cm$^{-1}$, which is specific for the Franck-Condon mechanism. One- and two-phonon Raman scattering both signal a JT orbital order in LaCoO$_3$ at least with short range correlations. This spectroscopic investigation supports and substantiates earlier thermodynamic experiments with respect to an intrinsic mixing of low spin and intermediate spin states at low temperatures in LaCoO$_3$.

Acknowledgments: This work was supported by INTAS Grant N 01-0278, NATO Collaborative Linkage Grant PST.CLG.977766 and DFG through SPP1073.

**Figures:**

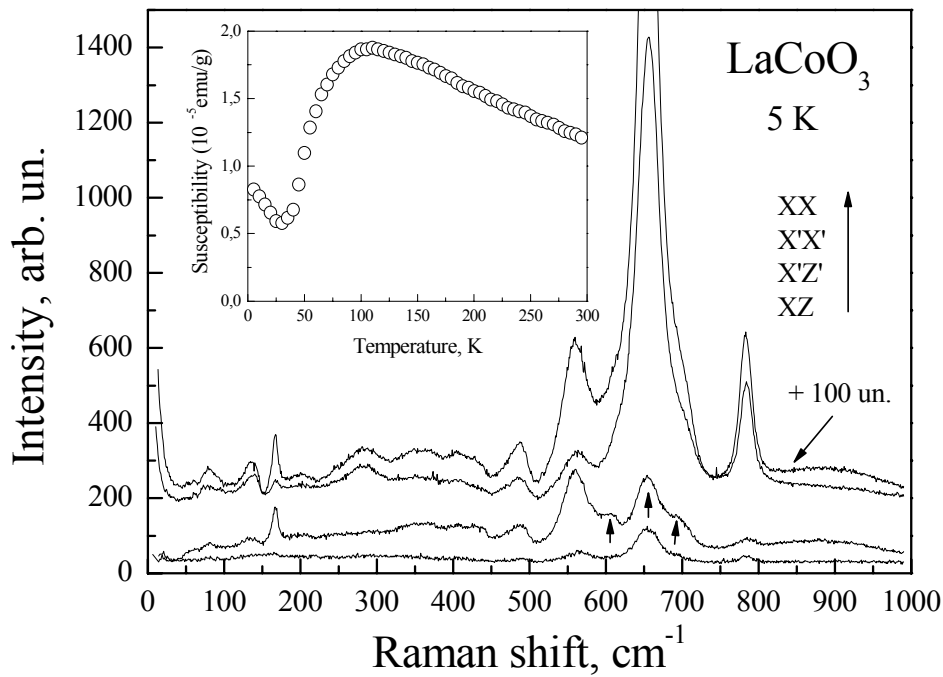

Fig. 1. Polarized Raman scattering spectra of single crystalline LaCoO$_3$ at 5 K. The arrows indicate the lines in the frequency region of stretching-like vibrations. The inset shows the temperature dependence of the magnetic susceptibility.

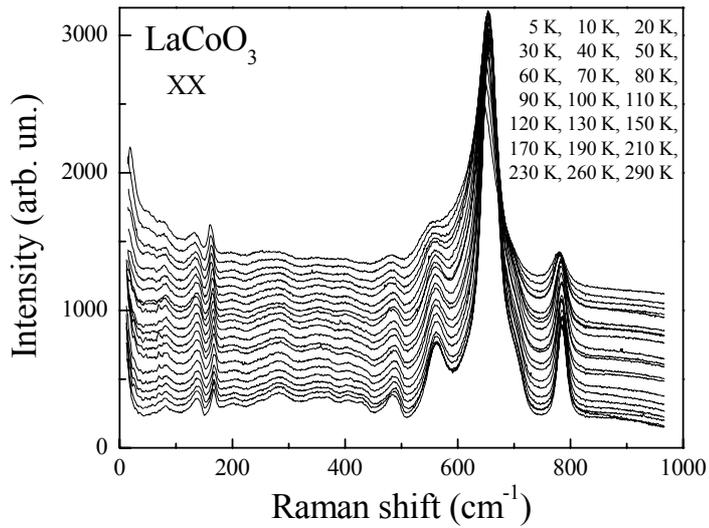

Fig. 2. Temperature dependence of the Raman spectra of single crystalline LaCoO$_3$ in *xx* scattering configuration. The spectra from 5 K (bottom) to 295 K (top) are shifted vertically for clarity.

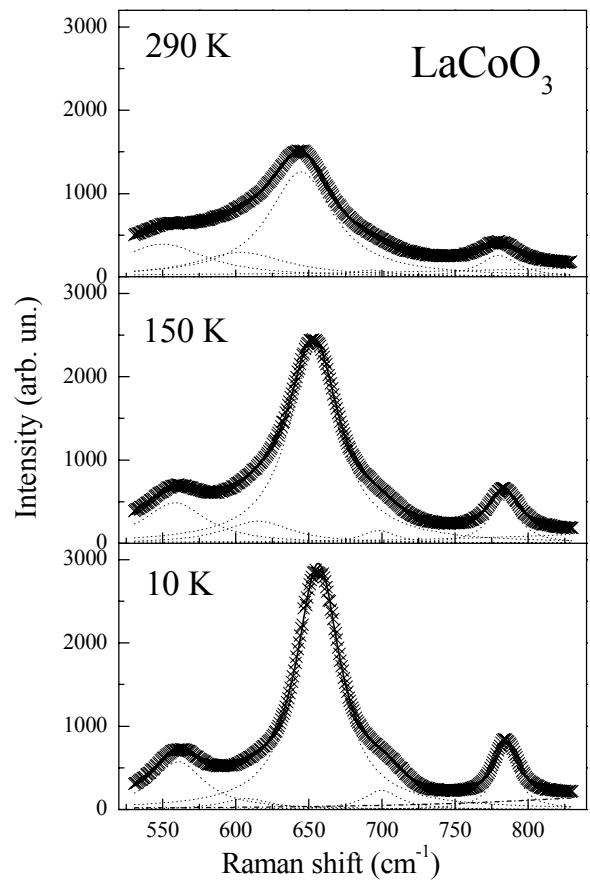

Fig. 3. A fit of the experimental Raman spectra to Lorentzian profiles.

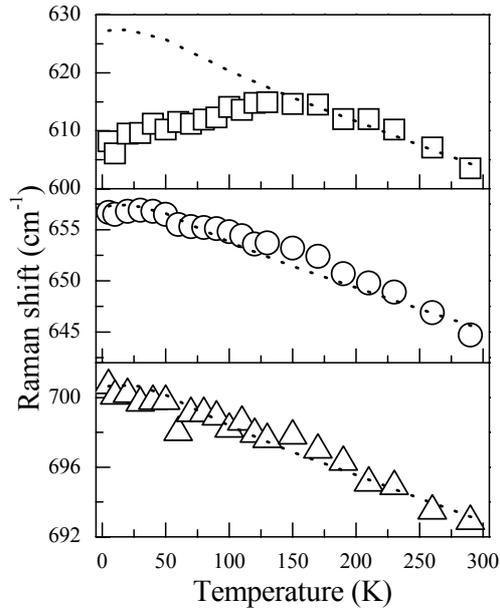

Fig. 4. Temperature dependence of the Raman shift for three phonon lines. The dotted lines are the estimated behavior of the phonon frequencies according to the Grüneisen law.

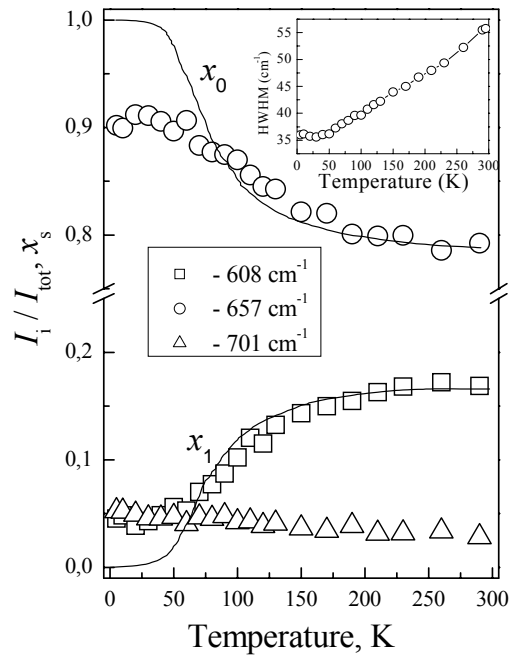

Fig. 5. Temperature dependence of the integrated intensity, $I_i$, of one-phonon lines at 608, 657, and 701 cm$^{-1}$ to their total integrated intensity, $I_{tot} = \sum_i I_i$. The solid lines present the scaled populations, $x_s$, of the Co$^{3+}$ ions in the LS ($x_0$) and IS ($x_1$) states. The inset shows the temperature dependence of the linewidth for the 657 cm$^{-1}$ line.

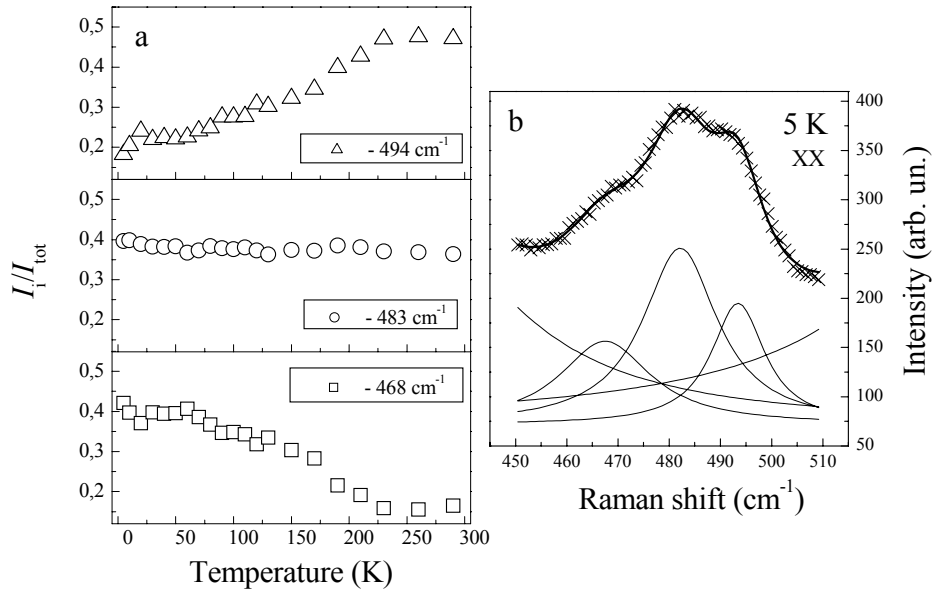

Fig. 6. (a)Temperature dependence of the integrated intensity, $I_i$, of one-phonon lines at 468, 483, and 494 cm$^{-1}$ normalized to their total integrated intensity and (b) an example of the spectra fitting in the frequency region of bending-like vibrations.

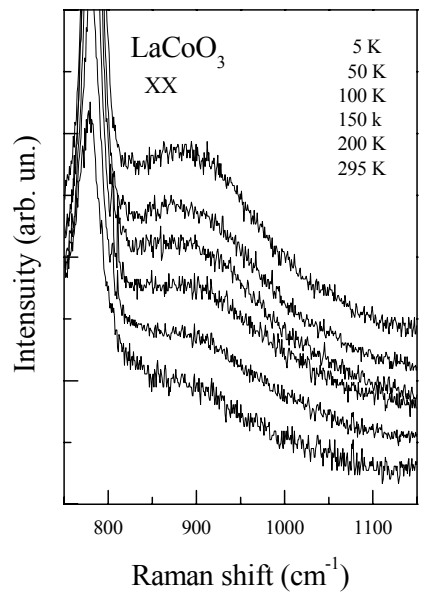

Fig. 7. Temperature dependence of the Raman spectra in the frequency region 750 – 1150 cm$^{-1}$ for the LaCoO$_3$ single crystal in *xx* scattering configuration. The spectra at different temperatures are shifted vertically for clarity.

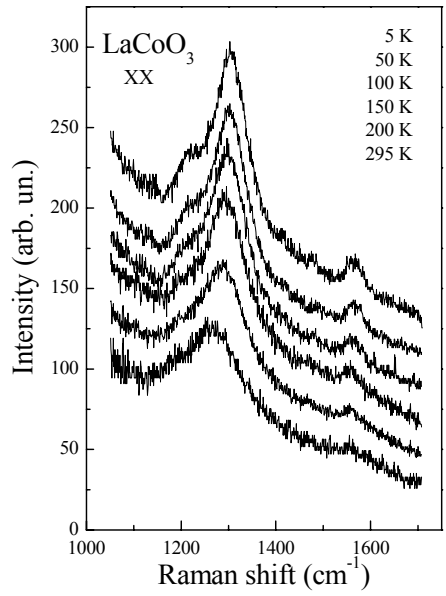

Fig. 8. Temperature dependence of the second-order Raman spectra in the LaCoO$_3$ single crystal. The spectra at different temperatures are shifted vertically for clarity.

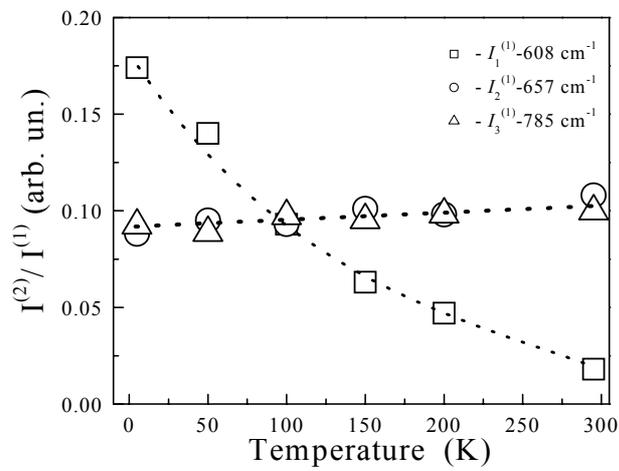

Fig. 9. The ratio of second- to first-order integrated Raman intensities ($I^{(2)}/I^{(1)}$). The dotted lines are guides to the eye.